\documentclass{epl}
\title{Synchronization Properties of Network Motifs}

\author{I. Lodato\inst{1}, S. Boccaletti\inst{2}, and V. Latora\inst{3}}

\institute{
  \inst{1} Scuola Superiore di Catania, Via S. Paolo 73, 95123 Catania, Italy\\
  \inst{2} CNR- Istituto dei Sistemi Complessi, Via Madonna del Piano, 10, 
                50019 Sesto Fiorentino (FI), Italy\\
  \inst{3} Dipartimento di Fisica e Astronomia, 
                    Universit\`a di Catania, and INFN, 
                Via S. Sofia 64, 95123 Catania, Italy }

\begin{document}
\newcommand{\FIXME}{{\bf FIXME}}
\newcommand{\real}{\mathrm{Real}}
\newcommand{\imag}{\mathrm{Imag}}

\maketitle

\begin{abstract}
We address the problem of understanding the variable abundance of 
3-node and 4-node subgraphs (motifs) in complex networks from a 
dynamical point of view.  
As a criterion in the determination of the functional significance 
of a $n$-node subgraph, we propose an analytic method to measure the  
stability of the synchronous state (SSS) the subgraph displays. 
We show that, for undirected graphs, the SSS is correlated with the 
relative abundance, while in directed graphs the correlation exists 
only for some specific motifs. 
\end{abstract}
\bigskip
\noindent

%
Recent empirical evidences indicate that complex networks, 
among other common properties, 
are characterized by the presence of various length cycles and 
specific motifs ~\cite{rev1,rev2,rev3}. 
A motif $M$ is a pattern of interconnections occurring either in a
undirected or in a directed graph $G$ at a number significantly
higher than in randomized versions of the graph, i.e. in graphs
with the same number of nodes and links (and eventually degree 
distribution) as the original one, 
but where the links are distributed at random. 
As a pattern of interconnections, is usually meant a small 
connected (undirected or directed) $n$-node graph $M$ which is a
subgraph of $G$. 
The concept of motifs was originally introduced by U. Alon and 
coworkers, who studied motifs in biological and non-biological 
networks \cite{alonNatGen02,alonScience02,alonPNAS03,
alonScience04,alonBioinformatics04}. 
The research of the significant
motifs in a graph $G$ is based on matching algorithms counting the
total number of occurrences of each $n$-node subgraph $M$ in the
original graph and in the randomized ones. The statistical
significance of $M$ is then described by the {\em $Z$-score},
defined as: 
$ Z_M=\frac{ \#_M - \langle \#_M^{rand} \rangle }
          { \sigma_{\#_M}^{rand}              } $, 
where $\#_M$ is the number of times the subgraph $M$ appears in
$G$, and $\langle \#_M^{rand} \rangle$ and $ \sigma_{\#_M}^{rand} $
are, respectively, the mean and standard deviation of the number of
appearances in the randomized network 
ensemble \cite{alonScience02,alonScience04}. 
The reasons of the variable frequency of different 
$n$-node subgraphs in a specific network are still poorly 
understood. 
There are at least two possible explanations. On the one hand,  
it is possile that certain constraints on the growth mechanism 
of a network as a whole determine which motifs become 
abundant \cite{sole,vazquez}. 
On the other hand, it is well known that the structure   
has important consequences on the network dynamics and functional 
robustness. So that a particular $n$-node graph can become 
overrepresented because, due to its structure, it possesses some 
relevant functional properties \cite{alonScience02}.  

In this letter, we address the question of network motifs 
in biological networks from a dynamic systems point of view. 
Naturally, a comprehensive analysis of the dynamics of networks  
is considerably more complicated than the corresponding analysis 
of their structure. This is due to the potentially complex functional 
dependencies between nodes, and to lack of knowledge of the specific 
interaction parameters.  
For such a reason, instead of modeling in details one particular biological network, 
we analyze the generic dynamic properties that arise from the topology 
of a $n$-node graph. In particular, we focus on the emergence of collective 
dynamic behaviors, such as {\em synchronization}, that is relevant in many 
biological systems, and we propose an analytic method to estimate  
the stability of the synchronous state (SSS) displayed by a $n$-node graph. 
We finally show that the SSS, potentially, can help explaining why 
certain network motifs are overrepresented in some real 
biological networks, while others are not.

%
We assume that the dynamics of a $n$-node motif  
$M$ can be represented as a system of $n$ ODE's: 
\begin{equation}
\label{ode}
    \begin{array}{l l}
    \dot {\bf x}_i = {\bf f}_i({\bf x}_1,{\bf x}_2,\ldots,{\bf x}_n)  & ~~~~~i = 1, \ldots, n
    \end{array}
\end{equation}
where ${\bf x}_i \in {R}^m$ is the $m$-dimensional vector
describing the state of node $i$ (for instance the concentration of 
molecule $i$ in a metabolic reaction, or the polarization state 
of neuron $i$ in a neural network), 
and ${\bf f}_i : {R}^{m \times n} \to {R}^m$ is the function representing 
the effects on ${\bf x}_i $ of all the nodes connected to $i$. In particular, we are 
neglecting the influence of the other nodes of the graph $G$ on the 
$n$-node motif $M$, and we are assuming that 
the ${\bf f}_i$'s do not contain an explicit dependence on time. 
The issue of the stability of the steady states of Eqs.~(\ref{ode}), i.e. the sets of values 
$({\bf x}^*_1, {\bf x}^*_2, \ldots, {\bf x}^*_n)$ such that 
$\dot{\bf x}^*_1=  \dot{\bf x}^*_2=  \ldots = \dot{\bf x}^*_n=0$, has been 
investigated in Ref.~\cite{prill}.  
Here we focus on the stability of the synchronized dynamics of Eqs.~(\ref{ode}),  
which can be treated analytically within the context of the so-called 
{\em Master Stability Function} approach \cite{pec98,fink00,bar02,rev3}. 
In particular, we restrict to the case in which the equations of motion 
can be written as: 
\begin{equation}\label{SCO1}
    \begin{array}{l l}
        \dot{\bf x}_i = {\bf F}_i ({\bf x}_i) + \sigma \sum_{j=1}^{n} 
{a}_{ij} [ {\bf H}_{ij}({\bf x}_j) -  {\bf H}_{ii}({\bf x}_i) ]
~~~~ i = 1, \ldots, n.
    \end{array}
\end{equation}
where ${\bf F}_i({\bf x}) : {R}^m \to {R}^m$ is the function 
governing the local dynamics of node $i$, 
${\bf H}_{ij} ({\bf x}_j) : {R}^m \to {R}^m $ 
describes the influence of node $j$ on node $i$,    
$\sigma>0$ is the coupling strength, and ${a}_{ij}$ are 
the elements of the $n \times n$ adjacency matrix of graph $M$.  
In the case of a undirected $n$-node motif $M$, ${a}_{ij}={a}_{ji}=1$  
iff there is an edge joining node $i$ and 
node $j$, and ${a}_{ij}={a}_{ji}=0$ otherwise. 
In the case of a directed motif, we assume  ${a}_{ij}=1$ iff  
there is a directed edge from node $j$ to node $i$, while 
${a}_{ij}=0$ otherwise. Equations (\ref{SCO1}) can be rewritten as: 
\begin{equation}\label{SCO2}
    \begin{array}{l l}
        \dot{\bf x}_i = {\bf F}_i ({\bf x}_i) 
- \sigma \sum_{j=1}^{n} {l}_{ij} {\bf H}_{ij}({\bf x}_j) ~~~~ i = 1, \ldots, n.
    \end{array}
\end{equation}
where ${l}_{ij} = \delta_{ij} (\sum_l a_{il}) -a_{ij}$  are 
the elements of a zero row-sum ($\sum_{j} l_{ij}  =0 \ \forall i$) 
$n \times n $ matrix $L$ with strictly positive diagonal terms 
($l_{ii} > 0 \ \forall i$). 
In the case of a undirected motif $M$,   
$L$ is symmetric and coincides with the standard {\em Laplacian matrix}  
of the graph $M$ \cite{rev3}.   
In the case of a directed graph, the off-diagonal 
elements $l_{ij}$ of $L$ are respectively equal to $-a_{ij}$, 
while the $i$-th diagonal entry 
is equal to the in-degree of node $i$, $k_i^{in} = \sum_l a_{il}$. 
In order to proceed with the analytic treatment, we make the
explicit assumption that the network is made of $n$ identical and 
identically coupled dynamical systems. 
This corresponds to take in Eqs.~(\ref{SCO1}) and Eqs.~(\ref{SCO2})  
${\bf F}_i({\bf x}_i) \equiv {\bf F}({\bf x}) ~ \forall i$, and 
${\bf H}_{ij}({\bf x}_j) \equiv {\bf H}({\bf x}) ~ \forall i,j$.  
This assumption and the fact that $L$ is zero-row sum, ensure the 
existence of an invariant set ${\bf x}_1 (t)= {\bf x}_2 (t)= \cdots  
= {\bf x}_n (t) \equiv {\bf x}_s (t)$, representing the complete  
synchronization manifold $\cal S$. 
The main idea, first proposed by Pecora and Carrol \cite{pec98}, is that 
the linear stability analysis of the synchronized state 
of Eqs.~(\ref{SCO2}) can be divided into a topological and a 
dynamical part \cite{pec98,bar02}. 
Since the coupling term of Eqs.~(\ref{SCO2}) vanishes exactly on $\cal S$, 
a necessary condition for the stability of the synchronous state is 
that the set of $(n-1)*m$ Lyapunov exponents corresponding to phase space 
directions transverse to the synchronization manifold 
are  entirely made of negative values. 
Considering, then, the $m \times n$ column vectors
$\mathbf{X}=(\mathbf{x}_{1},\mathbf{x}_{2},\ldots,\mathbf{x}_{n})^T$
and $\delta\mathbf{X}=(\delta \mathbf{x}_{1},\ldots, \delta
\mathbf{x}_{n})^T$, where 
$\delta {\bf x}_{i}(t)={\bf x}_{i}(t)-{\bf x}_s (t)$ 
is the deviation of the $i^{\tx th}$ vector state from the 
synchronization manifold, one gets the variational equation: 
\begin{eqnarray}
  \delta\dot{\mathbf{X}} & = &
  \left[{I}_{n}\otimes {\tx J}\mathbf{F}(\mathbf{x}_{s}) -
  \sigma
   {L} \otimes {\tx J}\mathbf{H}(\mathbf{x}_{s})\right]\delta\mathbf{X},
  \label{eq:var}
\end{eqnarray}
where $I_{n}$ is the $n \times n $ identity matrix, 
$\otimes$ stands for the direct product between matrices,
and ${\tx J}$ denotes the Jacobian operator. 
The first term in  Eq.~(\ref{eq:var}) is block diagonal 
with $m \times m$ blocks, while the second term can be treated 
by diagonalizing $L$.

%
We first concentrate on the case of {\bf undirected motifs}, 
i.e. on symmetric and thus diagonalizable laplacian $L$. 
Let $\lambda_i$ be the set of $n$ real eigenvalues of $L$ 
($ L {\bf v}_i = \lambda_{i} {\bf v}_i$, $i=1,\ldots,n$), 
and ${\bf v}_i$ the associated orthonormal eigenvectors 
($\mathbf{v}_{j}^{T}\cdot\mathbf{v}_{i}=\delta_{ij}$). 
If $L$ is symmetric, all its eigenvalues are real, 
and they can be ordered by size as:  
$0=\lambda_1 \leq \lambda_2 \leq \ldots \leq \lambda_n$.  
The arbitrary state $\delta\mathbf{X}$ can be written as
$\delta\mathbf{X}=\sum_{i=1}^{n}\mathbf{v}_{i}\otimes\zeta_{i}(t)$, where 
$\zeta_{i} \equiv ( \zeta_{1,i},\ldots,\zeta_{m,i} )$.
By substituting into Eq.~(\ref{eq:var}), and using the condition that 
the eigenvectors are linearly independent, one is finally left with 
a block diagonalized variational equation, with each of the $n$ blocks  
having the form of a variational equation for the coefficient  
$\zeta_{k}(t)$: 
\begin{eqnarray}
  \frac{d\zeta_{k}}{dt}  = \mathbf{K}_k \zeta_{k}, ~~~~ k=1,\ldots,n
  \label{diago}
\end{eqnarray}
where $\mathbf{K}_k =\left[{\tx J}\mathbf{F}(\mathbf{x}_{s}) - \sigma
    \lambda_{k}{\tx J}\mathbf{H}(\mathbf{x}_{s})\right]$ 
is the evolution kernel. Each equation in (\ref{diago})
corresponds to a set of $m$ conditional Lyapunov exponents 
along the eigenmode corresponding to the specific eigenvalue $\lambda_k$.
For $k=1$, $\lambda_1=0$, and we have the variational equation for the 
synchronized manifold ${\cal S}$. The $m$ corresponding 
conditional Lyapunov exponents equal those of the single uncoupled system 
$\dot{\bf x} = {\bf F}({\bf x})$, therefore no
conditions on them will be imposed 
(in principle, the synchronized state itself can well have positive 
Lyapunov exponents and be chaotic).

Notice that the Jacobian ${\tx J}\mathbf{F}(\mathbf{x}_{s})$ and 
${\tx J}\mathbf{H}(\mathbf{x}_{s})$ 
are the same for each block $k$, since they are evaluated on the 
synchronized state. Consequently, the form of each of the blocks in 
Eqs.~(\ref{diago}) is the same, with the only difference being in the 
multiplier $\lambda_k$. This leads one to replace $\sigma \lambda_k$ by 
$\nu$ in Eq.~(\ref{diago}), and to consider the generic $m$-dimensional
variational equation:
\begin{equation}
\label{diagonu}
\dot{\zeta} = \mathbf{K}_\nu \zeta = \left[ \tx{J}{\bf
F}({\bf x}_s)  - \nu \tx{J}{\bf H} ({\bf x}_s) \right]
\zeta,
\end{equation}
from which one can extract the set of $m$ conditional Lyapunov
exponents as a function of the real parameter $\nu \ge 0$. 
The parametrical behavior of the largest of such exponents, 
$\Lambda(\nu)$, is called {\it Master Stability Function} 
\cite{pec98,fink00,bar02}. 
In fact, given a coupling strength $\sigma$, one can locate the point 
$\sigma \lambda_k$ on the positive  $\nu$ axis, and the sign of 
$\Lambda$ at that point will reveal the stability of that eigenmode. 
If $\Lambda (\sigma \lambda_k) < 0 ~\forall k=2,...,n$, then the synchronous 
state is stable at the coupling strength $\sigma$. 

\begin{figure}[tb]
\centerline{\includegraphics*[scale=0.40]{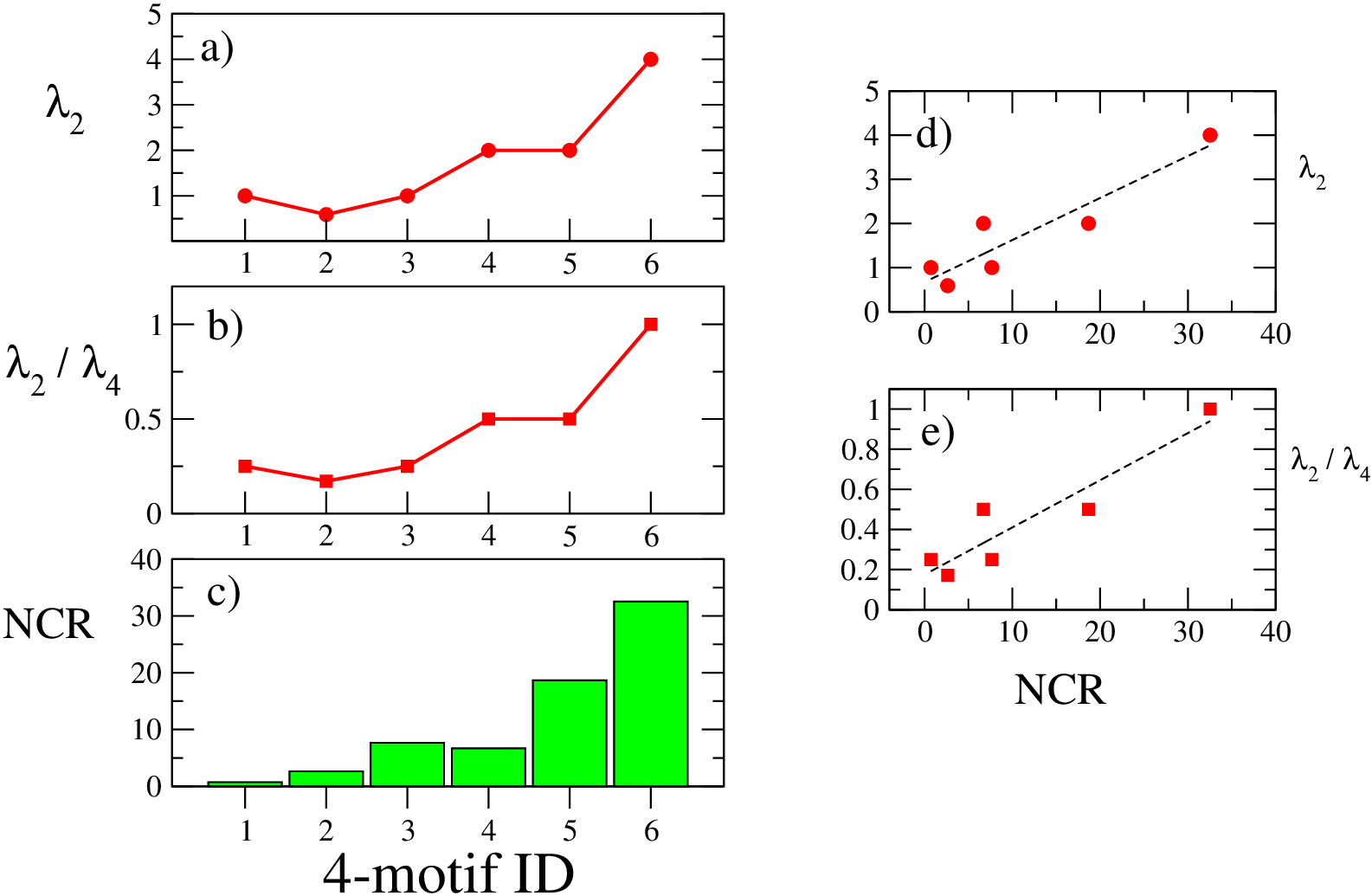}}
\caption{The value of SSS for each of the six 4-node undirected motifs are 
reported in panel a) for class II systems and in panel b) for class III, and   
compared with the natural conservation rates (NCR) in the yeast protein 
interaction network ~\cite{wuchty}. The motif identification number is the same as in Ref.~\cite{alonScience04}. In panel d) and e) we plot the values of SSS 
as function of the NCR (symbols), and the linear fittings obtained 
(dashed lines).} 
\label{figure_4}
\end{figure}

In order to evaluate whether the stability of the synchronous state is 
favoured by the topology in a given $n$-node graph more than in another, 
we adopt the following measures of stability. 
First, we assume that $\Lambda(\nu=0)> 0$, meaning that the uncoupled  
systems $\dot{\bf x} = {\bf F}({\bf x})$ support a chaotic 
dynamics. For $\nu>0$, there are three possible behaviors of 
$\Lambda(\nu)$,  
defining three possible classes for the choice of the functions 
${\bf F}({\bf x})$ and ${\bf H} ({\bf x})$.  
Case I (II) corresponds to a monotonically increasing (decreasing) 
$\Lambda(\nu)$. 
Case III admits negative values of $\Lambda(\nu)$ in 
the range $\nu_{c_1} < \nu < \nu_{c_2}$ (see Fig 5.1 of Ref.\cite{rev3}).  
For systems in class I, one can never stabilizes synchronization in  
any graph topology. In fact, for any $\sigma$ and any 
eigenvalues' distributions, the product 
$\sigma \lambda_k$ always leads to a positive maximum 
Lyapunov exponent, and therefore the synchronization manifold
$\cal S$ is always transversally unstable. 
Class II systems always admits synchronization for a large enough $\sigma$. 
In fact, given any eigenvalue distributions (any graph topology) 
it is sufficient to select $\sigma > \nu_c/ \lambda_2$ 
($\lambda_2 \neq 0$ in a connected graph \cite{fiedler})
to warrant that all transverse directions to $\cal S$ have associated negative 
Lyapunov exponents. The synchronous state 
will be stable for smaller values of $\sigma$ in a graph with a 
larger $\lambda_2$, so that $\lambda_2$ can be used as a measure of 
the stability of the synchronous state (SSS). 
For systems in class III, the stability condition is satisfied 
when $\sigma > \nu_{c_1}/ \lambda_2$ and $\sigma < \nu_{c_2}/ \lambda_N$, 
indicating that the more packed the
eigenvalues of $L$ are, the higher is the chance of having
all Lyapunov exponents into the stability range~\cite{bar02}. 
Consequentely, the ratio $\lambda_2 /\lambda_n$ 
can be used as a measure of SSS. 
Classes II and III include a large number of functions $F$,   
describing several relevant dynamical systems, as
the Lorenz and R\"ossler chaotic oscillators, and the Chua oscillator.  
It is important to notice that not only $F$, but also $H$ has a role 
in determining to which class a specific dynamical system belongs to. As an example, 
a nearest neighbor diffusive coupling on the R\"ossler chaotic 
system yields a class II (class III) Master Stability Function, when the
function $H$ extracts the second (the first) component of the vector
field \cite{hwang}. 
In Fig.~\ref{figure_4} (panel a and b) we report the two indices of SSS, 
namely $\lambda_2$ (class II) and $\lambda_2 /\lambda_4$ (class III), 
for the six 4-node undirected motifs. 
We observe a general increase in the SSS's as the number of the edges in 
the motif increases. Such an increase in SSS is in agreement with 
the decrease of the synchronization threshold observed numerically 
in the Kuramoto model by Moreno et al. \cite{moreno}. 
The two measures of SSS we propose are also in good agreement with the natural 
conservation ratio (NCR) for the same 4-node motifs in the 
the yeast protein interaction network reported in panel c).   
The NCR is a measure proposed in Ref.\cite{wuchty} to quantity 
the conservation of a given motif in the evolution across species, 
and is highly correlated to the motif Z-score.   
In panel d) and e) we show that SSS's and NCR are linearly correlated:   
we have obtained a correlation coefficient respectively equal 
to 0.94 and 0.93. 
This is an indication that motifs displaying an improved stability of 
cooperative activities (as synchronous states) are preserved across 
evolution with a higher probability.
%

%
We now turn our attention to {\bf directed motifs}. 
In a directed graph, the matrix $L$ is asymmetric and in general 
not always diagonalizable. Nevertheless, $L$ can be transformed into 
a Jordan canonical form, and it has been proven that the same condition 
valid for diagonalizable networks 
($\Lambda (\sigma \lambda_k) < 0 ~\forall k=2,...,n$) 
also applies to non-diagonalizable networks \cite{nishi}. 
In addition, the spectrum of $L$ is either real 
or made of pairs of complex conjugates. Because of the zero row-sum condition,  
$L$ always admits $\lambda_1=0$, and the other eigenvalues 
$\lambda_k=\lambda^R_k + i \lambda^I_k, \ k=2, \ldots, n$ 
(having non negative real parts according to the Gerschgorin's circle theorem
\cite{ger31}) can be ordered by increasing real part 
($0 \leq \lambda^R_2 \leq \ldots \leq \lambda^R_n$). 
Consequently, the parametric equation (\ref{diagonu}) has to be studied 
for complex values of the parameter $\nu=\nu^R + i \nu^I$. 
This yields a master stability function $\Lambda(\nu)$ as a surface over the complex 
plane $\nu$, that generalizes the plots for the case $\nu$ real. 
By calling $\cal R$ the region in the complex plane where $\Lambda(\nu)$ 
provides a negative Lyapunov exponent, the stability condition for the synchronous
state is that the set $\{ \sigma \lambda_k , k=2,\ldots,n \}$ be
entirely contained in $\cal R$ for a given $\sigma$. 
This is best accomplished for connection topologies that make 
${\lambda^R_2}$ as large as possible for class I systems, 
and for topologies that simultaneously make 
$\frac{\lambda^R_2}{\lambda^R_N}$ as large as possible 
and $\displaystyle\max_{k \geq 2} \{ \mid \lambda^I_k \mid \}$ as
small as possible, for class II systems.  

\begin{figure}[tb]
\centerline{\includegraphics*[scale=0.60]{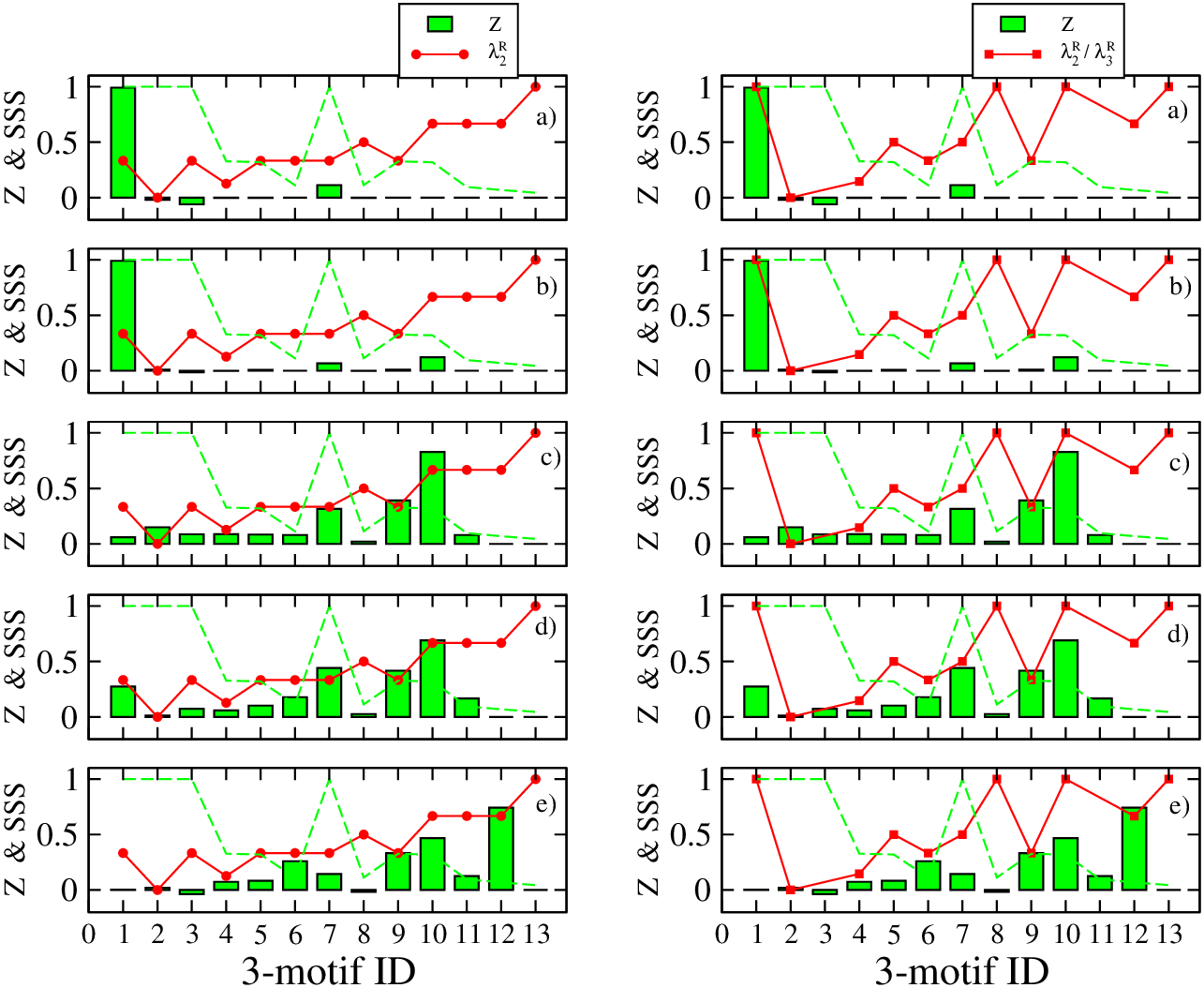}}
\caption{The SSS of each of the thirteen 3-node directed motifs is 
reported (continuous line) for class II (left panels), 
and class III systems (right panels). For class III system the SSS 
values has been normalized to the maximum value so to vary in the 
range [0,1].  
The SSS values are compared with the Z-score (hystograms) 
and with a measure of the stability of stationary states (dashed line) 
from Ref.\cite{prill}, in five different biological networks: 
the transcriptional regulatory networks of E.~coli (panels a) and 
S.~cerevisiae (panel b), 
the signal transduction knowledge environment (STKE) network  (panel c), 
the developmental transcriptional network of Drosophila melanogaster (panel d), and the neural connection map of C. elegans (panel e). } 
\label{figure_3d}
\end{figure}

\begin{figure}[tb]
\centerline{\includegraphics*[scale=0.30]{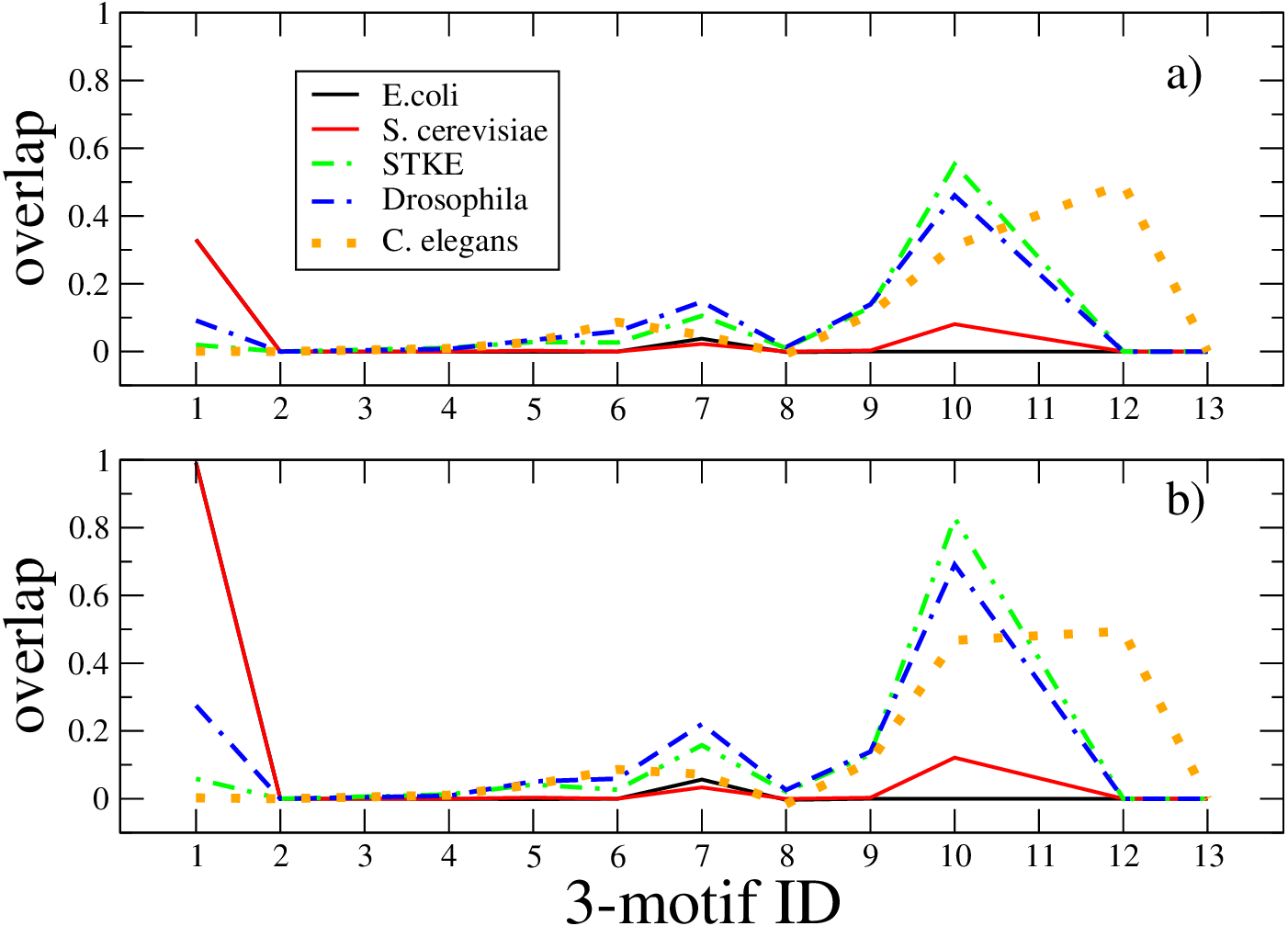}}
\caption{The overlap coefficients for each of the thirteen 
3-node directed motifs, and the five biological networks considered, 
are reported for class II (panel a) and class III systems (panel b).} 
\label{figure_3d_overlap}
\end{figure}

In Fig.~\ref{figure_3d}, we consider the thirteen 3-node directed motifs. 
Two of them, namely motifs \#3 and motif \#11 give rise to 
non-diagonalizable $L$. Motif \#8 is the only case where 
the eigenvalues are not real. 
In the left (right) panels we report $\lambda^R_2$ for class II systems 
($\lambda^R_2 /\lambda^R_3$ for class III systems). 
The SSS measures are compared with the Z-score profile obtained for five 
different real biological networks, and shown as hystograms in the figure.  
Both class I and class II systems exhibit an average increase of SSS 
as a function of the number of links in the motif. However, the 
overall agreement of the SSS and the Z-score profiles is not as good 
as in the case of undirected 4-motifs. Here, we have obtained rather 
small values (ranging from 0.1 to 0.3) of the correlation coefficient, 
with a better agreement found in the case of the STKE network (panels c), 
Drosophila (panels d) and C.elegans (panels e), rather than in  
the transcriptional regulatory networks (panels a and b). 
This might be due to the fact that synchronization processes are more 
important in neural systems than in other biological systems as 
transcriptional networks, especially the simplest ones 
(E.~coli and S.~cerevisiae). 
We have also reported in figure, as dashed lines, the measure of the stability 
of stationary states proposed by Prill et al. \cite{prill}. 
Such a measure seems to be better indicated for those 
systems where the stability of stationary states can be a more relevant 
dynamical quantity to investigate than the stability of synchronous states. 
Fig.~\ref{figure_3d} clearly indicates that in some motifs, 
SSS and Z-score are better correlated than in others. Hence, for each motif 
$M$, we have defined an {\em overlap coefficient} $O_M$ as: 
$O_M=SSS_M \times Z_M$. The maximum possible value $O_M=1$ indicates a perfect 
correlation between SSS and Z-score. The overlap coefficients obtained 
for the five studied systems are reported in Fig.~\ref{figure_3d_overlap}.  
For both class II and class III systems we have high values of the 
overlap for motifs: 1, 7, 10, 12.

Finally, we have considered the 199 4-node directed motifs. Here we 
report the results for three of the most statistically relevant motifs 
found in biological networks:  
the bifan, the biparallel and the feedback loop (see Ref.~\cite{alonScience02}). 
Such three motifs correspond all to cases in which $L$ can be diagonalized.   
The biparallel graph, that is abundant in the C. elegans and in 
transcriptional networks, has real eigenvalues and a relatively 
high value of SSSs: 
$\lambda^R_2=1$ and $\lambda^R_2 /\lambda^R_4=0.5$. 
The same is true for the 4-node feedback loop (also found abundant in 
electric circuits \cite{alonScience02}), having  $\lambda^R_2=1$, 
$\lambda^R_2 /\lambda^R_4=0.5$ and 
$\max_{k \geq 2} \{ \mid \lambda^{I}_k \mid \}=1$. 
Conversely, the bifan is not compatible with synchronization 
for any choice of ${\bf F}({\bf x})$ and ${\bf H} ({\bf x})$, 
and for any value of $\sigma$, since $\lambda_2= 0$ and we have assumed 
the case of networked chaotic systems ($\Lambda(\nu=0)> 0$). 
In fact, $\lambda^R_2 \neq 0$ iff the graph embeds an oriented
spanning tree, (i.e., there is a node from which all other nodes
can be reached by following directed links) \cite{wu,nishi} 
and this condition, that generalizes the notion of connectedness 
for undirected graphs \cite{fiedler} to directed graphs, 
is not valid in the case of the bifan.

We warmly thank R.J. Prill and A. Levchenko for having provided us 
with their results on the stability of stationary states,  
and G. Russo for useful comments.  
S.B. acknowledges the Yeshaya Horowitz Association through the 
Center for Complexity Science.

\end{document}